\documentclass[aps,prl,twocolumn,epsfig,showpacs]{revtex4}
\usepackage[centertags]{amsmath}
\usepackage{amssymb}
\usepackage{amsthm}
\usepackage{graphicx,subfigure}
\usepackage{epsfig}

\begin{document}

\title{System-reservoir theory with anharmonic baths: a perturbative approach}

\author{Chitrak Bhadra and Dhruba Banerjee}


\affiliation{Department of Physics, Jadavpur University, Kolkata, India}

\begin{abstract}
\noindent In this paper we present a study of a general system coupled to a reservoir (the words ``bath" and ``reservoir" will be used interchangeably) consisting of nonlinear oscillators, based on perturbation theory at the classical level. We extend the standard Zwanzig approach of elimination of bath degrees of freedom order by order in perturbation. We observe that the fluctuation dissipation relation (FDR) in its standard form for harmonic baths gets modified due to the nonlinearity and this is manifested through higher powers of $k_BT$ in the expression for two-time noise correlation. As an aside, we also observe that the first moment of the noise arising from a nonlinear bath can be non-zero, even in absence of any external drive, if the reservoir potential is asymmetric with respect to one of its minima, about which one builds up the perturbation theory.
\end{abstract}

\maketitle


This paper deals with the noise properties of a reservoir of classical nonlinear oscillators. The method of study follows the time-honoured paradigm of system-reservoir interaction based on the Hamilton's equation of motion for the system that eventually takes the shape of a generalized Langevin equation. The strategy of capturing a wide range of physical and chemical phenomena through the paradigm of system-reservoir interaction has a very rich history and hence the literature is colossal \cite{zwabook,zwa73,lind1,lind2,hanggi1,hanggi2,mazo,feyn,weiss,legg}. In particular, the kind of system-reservoir theory that we shall deal with in this paper follows the path of Zwanzig \cite{zwabook,zwa73}: one starts with a microscopic system-reservoir Hamiltonian, inclusive of all kinds of coupling among the system and reservoir oscillators, and then eliminates the reservoir degrees of freedom to arrive at a generalized Langevin equation that summarizes the structures of damping and forcing which the reservoir imparts on the system in tandem. While the damping can be identified with the strength of the couplings by which the system couples with the individual reservoir oscillators, the noise consists of some nontrivial combinations of the initial values of the dynamical variables pertaining to the reservoir. In the structures of both these quantities the coupling term plays a very important role. All this culminates in a fluctuation-dissipation relation (FDR) that equates the two time noise correlations with the damping \cite{kubo}.

Studies relating to system-reservoir interaction bank heavily on the fact that the reservoir consists of a huge number of non-interacting harmonic oscillators, which are linearly or nonlinearly coupled to the system. The system itself can be under the influence of some external potential and that can be nonlinear also. Though harmonic approximation and linearization always give a meaningful picture at the leading order of the dynamics, to know the effects of the nonlinearities associated therein -- either with a coupling that is nonlinear in the system coordinates and linear in those of the reservoir or vice versa -- one has to make a perturbative approach both at the classical and the quantum levels \cite{hu,lind2,legg}. The great virtue of such approaches is that, so long as the bath Hamiltonian remains harmonic in its oscillators, the nature of the coupling -- linear or nonlinear -- does  not violate the aforementioned structure of the FDR \cite{zwabook,zwa73,lind1,lind2}.

At this point it is relevant to ask, what happens to the FDR when the reservoir oscillators are themselves nonlinear. At first it may seem like unnecessary pedantry. But adding nonlinearities to the bath oscillators makes things much more complicated than usual and the results for the first and second moments of the noise differ from the cases alluded to in the previous paragraphs. For example, even if there is no external drive on the system or the reservoir, the first moment of the noise, or the stochastic term of the generalized Langevin equation, does not vanish in general. Starting from Einstein's theory of Brownian motion to all forms of undriven stochastic processes relevant to physical or chemical systems, all have the vanishing of the first moment of the noise as a basic requirement for the stochasticity of the dynamics to make any sense. But if there be an asymmetry in the reservoir potential (about some chosen origin which is some minimum of the potential), then this is no longer true. The reservoir potential, which we take to be anharmonic, itself is symmetric in structure but for building a perturbation theory one has to start from a certain minima, and with respect to this particular minimum the potential is asymmetric. It turns out that to first order in a perturbation parameter (here we have called it $\lambda$) that connects to the coupling term of the Hamiltonian, the first moment of the noise is zero. However, in the next higher order (here we have called that order as $O(\lambda\epsilon)$, where $\epsilon$ is a parameter that connects to the nonlinearity of the bath Hamiltonian) there is a temperature-dependent non-zero contribution arising from the aforementioned asymmetry in the reservoir potential. This fact is summarized in Eq.(\ref{eq3.20}) below. But, if the reservoir potential is symmetric, then the first moment of the noise vanishes as usual.

The two-time correlation of the noise also reveals new features at higher orders. It turns out that the structure of the FDR is robust so long as one is confined to calculations within order $O(\lambda^2)$ \cite{lind2,weiss}, which is the lowest relevant order for two time noise correlations. The next higher order, i.e., $O(\lambda^2\epsilon)$, where signatures of the nonlinearity of the bath Hamiltonian are non-negligible, yields non-zero terms that contain higher powers of $k_BT$, as opposed to a single power of $k_BT$ associated with the damping as is usually seen in FDRs which emerge from microscopic system-reservoir models. These results are summarized in Eqs.(\ref{eq3.42}) and (\ref{eq3.43}) below.

The paper is organized as follows. In Section I we construct the full microscopic Hamiltonian for a double-well bath. For a perturbation theory on a bound potential to make sense, the zeroth order solution must be harmonic. This requires some algebraic rearrangements on the reservoir Hamiltonian that makes it asymmetric about a shifted origin. The other terms in the Hamiltonian are those of the system, the coupling term and a counter-term discussed below. In Section II we construct the generalized Langevin equation to be satisfied by the system coordinate and this allows us to identify the microscopic structures of the damping and noise terms, both arising from the bath. Section III has been devoted to detailed calculation of the moments of the noise term that lead to conclusions summarized in the previous paragraph. In Section IV the case of a symmetric quartic bath potential has been considered the results of which follow very easily from the general expressions relating to the double-well bath considered exhaustively in the preceding three Sections. In Section V the paper has been concluded.

\section{I. The double-well Hamiltonian}

\noindent In this Section we introduce the full Hamiltonian for the system and reservoir. Generally a double well oscillator with the central maximum at the origin is written as $U(q) = -aq^2 + bq^4$, where $a$ and $b$ are positive constants and $q$ denotes the coordinate of a reservoir oscillator. But to effect a perturbative treatment, we always prefer the unperturbed state to be one of a simple harmonic oscillation. Therefore, at the very beginning, we rewrite the above potential by shifting the origin to the bottom of one of the wells, say, the left well. This shifting renders the potential asymmetric about this particular chosen minimum, albeit the potential itself is symmetric in structure. This leads to the expression for the potential of the $\mu$-th oscillator in the reservoir as

\begin{eqnarray}
U_{\mu} (q_{\mu}) &=& \frac{{a_{\mu}}^2}{2b_{\mu}} - a_{\mu} \left( q_{\mu} - \sqrt{\frac{a_{\mu}}{2b_{\mu}}} \right)^2
\nonumber \\
&+& b_{\mu} \left( q_{\mu} - \sqrt{\frac{a_{\mu}}{2b_{\mu}}} \right)^4.
\label{eq1.1}
\end{eqnarray}

\noindent
As discussed earlier, we remark here that this shifting of the origin breaks the symmetric structure of a double-well potential, the signature of which will appear in our calculation of the first moment of the noise later on.

\noindent The system-reservoir Hamiltonian should consist of the following components:

\begin{equation}
H = H_S + H_R + H_C + H_{CT}
\label{eq1.2}
\end{equation}

\noindent where $H_S$ stands for the Hamiltonian of the system -- here a particle of mass $M$ subjected to some general potential energy $V(x)$ -- and is hence given by

\begin{equation}
H_S = \frac{p^2}{2M} + V(x).
\label{eq1.3}
\end{equation}

\noindent
The $H_R$ term of Eq.(\ref{eq1.2}) stands for the reservoir Hamiltonian and consists of harmonic terms, and associated with them are the anharmonic terms which have to be treated perturbatively. In the form for $U_{\mu}$ in Eq.(\ref{eq1.1}) we do not have a positive quadratic term that might lead to a simple harmonic oscillation. Therefore, using Eq.(\ref{eq1.1}), we write the reservoir Hamiltonian as

\begin{eqnarray}
H_R &=& \sum_{\mu} \left[ \left\{\frac{{p_{\mu}}^2}{2m_{\mu}} + \frac{1}{2}m_{\mu} {\omega_{\mu}}^2 {q_{\mu}}^2 \right\} \right.
\nonumber \\
&+& \left. \epsilon \left\{ U_{\mu} (q_{\mu}) -  \frac{1}{2}m_{\mu} {\omega_{\mu}}^2 {q_{\mu}}^2 \right\} \right],
\label{eq1.4}
\end{eqnarray}

\noindent where we have brought a harmonic term in the zeroth order and nullified it in the 1st order of a perturbation parameter $\epsilon$. This is exact if $\epsilon = 1$, but otherwise, at the cost of an allowable distortion in the original Hamiltonian, this procedure paves the way for a perturbative treatment. This technique of bringing a harmonic term in the zeroth order and subsequently cancelling it in the next order has been successfully used in studying nonlinear dynamical systems in various dimensions \cite{jkb1,jkb2,jkb3} -- systems which by virtue of their structure, do not appear to be conducive for any standard perturbative treatment.

\noindent The term $H_C$ in Eq.(\ref{eq1.2}) represents the coupling between the system and the reservoir. In this work we take this coupling to be linear in both system and reservoir coordinates, with the coupling constant for the $\mu$-th oscillator represented as $C_{\mu}$. Thus

\begin{equation}
H_C = \lambda C_{\mu} q_{\mu}(t) x(t)
\label{eq1.5}
\end{equation}

\noindent where, it should be noted, we use a different perturbation parameter $\lambda$, keeping in mind the fact that, generally, the degree of smallness associated with the nonlinearity in the reservoir coordinates can be different from the degree of smallness by which the system and reservoir couple.

\noindent The final term in Eq.(\ref{eq1.2}), $H_{CT}$, is the counter-term that prevents the reservoir modes from modifying (or renormalizing) the system potential $V(x)$ \cite{weiss}. In the present scenario this is required to be a term second order in $\lambda$ and is given by

\begin{equation}
H_{CT} = \lambda^2 .\frac{{C_{\mu}}^2}{2m_{\mu} {\omega_{\mu}}^2}. x^2(t).
\label{eq1.6}
\end{equation}

\noindent Having thus formed the full Hamiltonian we now follow the standard prescription of deriving the (Langevin) equation of motion for the system coordinate $x$, by eliminating the bath degrees of freedom perturbatively. This will lead us to explicit expressions of the damping and the noise -- both arising from the reservoir -- and hence the relation between them, namely, the fluctuation-dissipation relation (FDR).

\section{II. The Generalized Langevin equation}

\noindent Differentiating the full Hamiltonian of Eq.(\ref{eq1.2}) with respect to $x$ we have

\begin{eqnarray}
\dot{p} &=& -\frac{\partial H}{\partial x}
\nonumber \\
&=& -V'(x) - \lambda\sum_{\mu} C_{\mu} q_{\mu}(t)
\nonumber \\
&+& \lambda^2 \sum_{\mu} \frac{{C_{\mu}}^2}{m_{\mu} {\omega_{\mu}}^2}. x(t)
\label{eq2.1}
\end{eqnarray}

\noindent and differentiating the same with respect to $p$ and combining that with the above equation yields the equation of motion for the system

\begin{equation}
M\ddot{x} + V'(x) = -\lambda \sum_{\mu} C_{\mu} q_{\mu}(t)
+ \lambda^2 \sum_{\mu} \frac{{C_{\mu}}^2}{m_{\mu} {\omega_{\mu}}^2}. x(t)
\label{eq2.2}
\end{equation}

\noindent from whose right hand side the reservoir coordinates have to be eliminated. The second term on the right hand side of Eq.(\ref{eq2.2}), arising from the counter-term of Eq.(\ref{eq1.6}), will eventually get cancelled. Similarly the equation of motion for the $\mu$-th reservoir mode is obtained as

\begin{eqnarray}
\ddot{q}_{\mu} + {\omega_{\mu}}^2 q_{\mu} &=& - \lambda \frac{C_{\mu}}{m_{\mu}} x(t) +
\epsilon \left[ \frac{2a_{\mu}}{m_{\mu}} \left( q_{\mu} - \sqrt{\frac{a_{\mu}}{2b_{\mu}}} \right)
\right. \nonumber \\
&-& \left. \frac{4b_{\mu}}{m_{\mu}} \left( q_{\mu} - \sqrt{\frac{a_{\mu}}{2b_{\mu}}} \right)^3
+ {\omega_{\mu}}^2 q_{\mu} \right].
\label{eq2.3}
\end{eqnarray}

\noindent
To solve this equation we now need to invoke a perturbation expansion written as

\begin{eqnarray}
q_{\mu} (t) &=& q_{\mu}^{(0)} (t) + \lambda q_{\mu\lambda}^{(1)} (t) + \epsilon q_{\mu\epsilon}^{(1)} (t)
\nonumber \\
&+& \lambda\epsilon q_{\mu\lambda\epsilon}^{(2)} (t)
+ \lambda^2 q_{\mu\lambda^2}^{(2)} (t)
+ \epsilon^2 q_{\mu\epsilon^2}^{(2)} (t)
+ \ldots.
\label{eq2.4}
\end{eqnarray}

\noindent
In this paper we shall confine ourselves to calculations which involve coordinate corrections only upto first-order in $\lambda$ and $\epsilon$. The $0$-th order equation that emerge by putting the above perturbative expansion in Eq.(\ref{eq2.3}) is

\begin{equation}
\ddot{q}^{(0)}_{\mu}(t) + {\omega_{\mu}}^2 q^{(0)}_{\mu}(t) = 0
\label{eq2.5}
\end{equation}

\noindent which being the equation of a simple harmonic oscillator yields the solution

\begin{equation}
q^{(0)}_{\mu} (t) = q_{\mu} (0) \cos \omega_{\mu}t + \frac{p_{\mu} (0)}{m_{\mu} \omega_{\mu}} \sin \omega_{\mu}t
\label{eq2.6}
\end{equation}

\noindent where $q_{\mu} (0)$ and $p_{\mu} (0)$ are initial values of coordinate and momentum of the $\mu$-th reservoir oscillator. In what follows, we have to use these two quantities so frequently that, for brevity, we shall henceforth omit the argument $(0)$ from them and shall use the notation

\begin{eqnarray}
Q_{\mu} &\equiv& q_{\mu} (0)
\label{eq2.7} \\
P_{\mu} &\equiv& p_{\mu} (0).
\label{eq2.8}
\end{eqnarray}

\noindent The first order equation in the perturbation parameter $\epsilon$, after some algebraic simplification, takes the form

\begin{eqnarray}
\ddot{q}_{\mu\epsilon}^{(1)} (t) + {\omega_{\mu}}^2 q_{\mu\epsilon}^{(1)} (t) &=&
{\Omega_{\mu}}^2 q^{(0)}_{\mu} (t)
\nonumber \\
&+& \frac{6}{m_{\mu}} \sqrt{2a_{\mu} b_{\mu}}~ [q^{(0)}_{\mu} (t)]^2
\nonumber \\
&-& \frac{4b_{\mu}}{m_{\mu}}~ [q^{(0)}_{\mu} (t)]^3,
\label{eq2.9}
\end{eqnarray}

\noindent where we have introduced the quantity $\Omega_{\mu}$ having the dimension of frequency as

\begin{equation}
\Omega_{\mu}  = \left[{\omega_{\mu}}^2 - \frac{4a_{\mu}}{m_{\mu}}\right]^{1/2}.
\label{eq2.10}
\end{equation}

\noindent The solution of this equation is obtained by repeated use of the right hand side of Eq.(\ref{eq2.6}). Thus,

\begin{eqnarray}
q_{\mu\epsilon}^{(1)} (t) &=& K_0 + K_{1s} \sin \omega_{\mu}(t) + K_{1c} \cos \omega_{\mu}(t)
\nonumber \\
&-& K_{2s} \sin 2\omega_{\mu}(t) - K_{2c} \cos 2\omega_{\mu}(t)
\nonumber \\
&-& K_{3s} \sin 3\omega_{\mu}(t) - K_{3c} \cos 3\omega_{\mu}(t).
\label{eq2.11}
\end{eqnarray}

\noindent The seven coefficients represented as different subscripts of $K$ are nontrivial combinations of $Q_{\mu}$ and $P_{\mu}$ and hence, these will be important for the calculation of noise correlations. We enlist them here:

\begin{eqnarray}
{\omega_{\mu}}^2 K_0 &=& \frac{3}{m_{\mu}} \sqrt{2a_{\mu} b_{\mu}}
\left[ {Q_{\mu}}^2 + \left\{ \frac{P_{\mu}}{m_{\mu} \omega_{\mu}} \right\}^2 \right]
\label{eq2.12} \\
4{\omega_{\mu}}^2 K_{1s} &=& {\Omega_{\mu}}^2 ~ \frac{P_{\mu}}{m_{\mu} \omega_{\mu}}
\nonumber \\
&-& \frac{3b_{\mu}}{m_{\mu}} \left[  \left\{ \frac{P_{\mu}}{m_{\mu} \omega_{\mu}} \right\}^3
+ \frac{ {Q_{\mu}}^2 P_{\mu}}{m_{\mu} \omega_{\mu}} \right]
\label{eq2.13} \\
4{\omega_{\mu}}^2 K_{1c} &=& {\Omega_{\mu}}^2 Q_{\mu}
\nonumber \\
&-& \frac{3b_{\mu}}{m_{\mu}} \left[ {Q_{\mu}}^3 + Q_{\mu} \left\{ \frac{P_{\mu}}{m_{\mu} \omega_{\mu}} \right\}^2 \right]
\label{eq2.14} \\
3{\omega_{\mu}}^2 K_{2s} &=& \frac{6\sqrt{2a_{\mu}b_{\mu}}}{m_{\mu}}
\frac{Q_{\mu} P_{\mu}}{m_{\mu} \omega_{\mu}}
\label{2.15} \\
3{\omega_{\mu}}^2 K_{2c} &=& \frac{3\sqrt{2a_{\mu}b_{\mu}}}{m_{\mu}}
\left[ {Q_{\mu}}^2 - \left\{ \frac{P_{\mu}}{m_{\mu} \omega_{\mu}} \right\}^2 \right]
\label{eq2.16} \\
8{\omega_{\mu}}^2 K_{3s} &=& \frac{b_{\mu}}{m_{\mu}}
\left[ \left\{ \frac{P_{\mu}}{m_{\mu} \omega_{\mu}} \right\}^3 - \frac{3{Q_{\mu}}^2 P_{\mu}}{m_{\mu} \omega_{\mu}} \right]
\label{eq2.17} \\
8{\omega_{\mu}}^2 K_{3c} &=& \frac{b_{\mu}}{m_{\mu}}
\left[ 3 Q_{\mu} \left\{ \frac{P_{\mu}}{m_{\mu} \omega_{\mu}} \right\}^2 - {Q_{\mu}}^3 \right]
\label{eq2.18}
\end{eqnarray}

\noindent Having thus obtained the solution of Eq.(\ref{eq2.9}) in full detail, we now go for the other perturbation parameter $\lambda$. The equation which is easily obtained from Eq.(\ref{eq2.3}) is

\begin{equation}
\ddot{q}_{\mu\lambda}^{(1)} (t) + {\omega_{\mu}}^2 q_{\mu\lambda}^{(1)} (t) =
-\frac{C_{\mu}}{m_{\mu}} x_0(t).
\label{eq2.19}
\end{equation}

\noindent Here $x_0(t)$ is the $0$-the order solution of Eq.(\ref{eq2.2}) emerging out of a perturbative expansion of the system coordinate $x$ similar to that done in Eq.(\ref{eq2.4}). The solution of Eq.(\ref{eq2.19}) is obtained by direct integration as

\begin{eqnarray}
q_{\mu\lambda}^{(1)} (t) &=& -\frac{C_{\mu}}{m_{\mu} {\omega_{\mu}}^2} x_0(t)
+ \frac{C_{\mu}}{m_{\mu} {\omega_{\mu}}^2} \left[ x(0) \cos \omega_{\mu}t \right.
\nonumber \\
&+& \left. \int_0^{\infty} dt' \dot{x}_0(t') \cos \omega_{\mu} (t - t') \right].
\label{eq2.20}
\end{eqnarray}

\noindent When this solution along with Eqs.(\ref{eq2.6}) and (\ref{eq2.11}) are substituted for $q_{\mu}(t)$ on the right hand side of Eq.(\ref{eq2.2}) in accordance with the perturbation expansion of Eq.(\ref{eq2.4}), then we see that the first term on the right hand side of Eq.(\ref{eq2.20}) gives a $\lambda^2$ contribution which cancels the second term on the right hand side of Eq.(\ref{eq2.2}). It should be recalled that the source of this term was the counter-term $H_{CT}$ we introduced in Eq.(\ref{eq1.6}), and hence its role is now over.

\noindent Finally, bringing all terms in one place, we have the generalized Langevin equation for the system coordinate as

\begin{equation}
M\ddot{x} + V'(x) + \int_0^t \gamma(t - t') \dot{x}_0(t) = \Gamma(t)
\label{eq2.21}
\end{equation}

\noindent where the damping term is given by

\begin{equation}
\gamma(t-t') = \lambda^2 \sum_{\mu} \frac{{C_{\mu}}^2}{m_{\mu} {\omega_{\mu}}^2} \cos \omega_{\mu}(t - t').
\label{eq2.22}
\end{equation}

\noindent Although, formalistically, we had implicitly invoked a perturbative expansion in the system coordinate $x$ as well in order to deal with the perturbation parameters sitting on the right hand side of Eq.(\ref{eq2.2}), within the damping kernel on the left hand side of Eq.(\ref{eq2.21}) we can replace $x_0(t)$ by $x(t)$ itself. This is because the relaxation time scale of the system -- which is a Brownian particle -- being much slower than the reservoir time scale which is of the order of $1/\gamma$, the difference between $x_0$ and $x$ within the damping kernel is not significant. To put it otherwise, the observational time scale over which one registers a noticeable change in the dynamics of $x$ as a result of the perturbation is much longer than the relaxation time of the reservoir degrees of freedom. Therefore we can rewrite Eq.(\ref{eq2.21}) as

\begin{equation}
M\ddot{x} + V'(x) + \int_0^t \gamma(t - t') \dot{x}(t) = \Gamma(t).
\label{eq2.23}
\end{equation}

\noindent Choosing the initial condition for the system as $x(0) = 0$, which we can always do, the noise term can be written as

\begin{equation}
\Gamma(t) = -\lambda \Gamma_{\lambda}(t) - \lambda\epsilon \Gamma_{\lambda\epsilon}(t)
\label{eq2.24}
\end{equation}

\noindent where the first order and the second order components of the noise are respectively given by

\begin{equation}
\Gamma_{\lambda}(t) = \sum_{\mu} C_{\mu} \left[ Q_{\mu} \cos \omega_{\mu}t
+ \frac{P_{\mu}}{m_{\mu} \omega_{\mu}} \sin \omega_{\mu}t \right]
\label{eq2.25}
\end{equation}

\noindent and

\begin{eqnarray}
\Gamma_{\lambda\epsilon}(t) &=& \sum_{\mu} C_{\mu}
[K_0 + K_{1s} \sin \omega_{\mu}(t) + K_{1c} \cos \omega_{\mu}(t)
\nonumber \\
&-& K_{2s} \sin 2\omega_{\mu}(t) - K_{2c} \cos 2\omega_{\mu}(t)
\nonumber \\
&-& K_{3s} \sin 3\omega_{\mu}(t) - K_{3c} \cos 3\omega_{\mu}(t)].
\label{eq2.26}
\end{eqnarray}

\noindent Our next objective is to calculate the first and second moments of this noise, which consists of several higher harmonic terms arising due to the nonlinearity present in the reservoir Hamiltonian. One relevant question to ask at this point is, do these higher harmonics leave any footprints in the fluctuation-dissipation relation ? We shall see that apart from modifying the bath spectrum by contributing some nonlinear powers of $\omega_{\mu}$ associated with higher powers of $k_BT$, the higher harmonics do not remain explicitly present in the FDR. Such a picture is far from being obvious at this stage of the discussion.

\section{III. Noise Correlations}

\noindent Due to some calculational heaviness of the following material we divide this Section into three subsections in an endeavour to enhance readability. Most of the integrals involve straightforward Gaussian integrations which can be evaluated by elementary Gamma-functions. Therefore to avoid making the manuscript awkwardly long, we show only the main steps here. The key results of this Section are Eq.(\ref{eq3.20}) and Eqs.(\ref{eq3.42}). In the former, the first moment of the noise has been evaluated, which, in the present set-up, can be non-zero. In the latter, the two time correlation of the noise has been calculated and this leads to a perturbatively corrected form of the FDR, the correcton coming in some nonlinear power of $k_BT$ upto the order of perturbation probed.

\subsection{1. The canonical partition function}

\noindent The calculation of the first and second moments of the noise will be done, as usual, on a canonical distribution of the initial values of all the reservoir degrees of freedom, i.e.,

\begin{equation}
\mathbb{P} (\{Q_{\mu}\}, \{P_{\mu}\}) = Z^{-1} \exp[-\beta(H_0 + \epsilon H_{1\epsilon})]
\label{eq3.1}
\end{equation}

\noindent where $Z$ is the canonical partition function with $\beta = k_BT$ and we identify $H_0$ and $H_{1\epsilon}$ from the expression of $H_R$ in Eq.(\ref{eq1.4}) as

\begin{equation}
H_0 = \sum_{\mu} \left[ \frac{{P_{\mu}}^2}{2m_{\mu}} + \frac{1}{2}m_{\mu} {\omega_{\mu}}^2 {Q_{\mu}}^2 \right]
\label{eq3.2}
\end{equation}

\noindent and, using Eq.(\ref{eq1.1}) and the perturbative part of Eq.(\ref{eq1.4}),

\begin{eqnarray}
H_{1\epsilon} &=& \sum_{\mu} \left[\frac{{a_{\mu}}^2}{2b_{\mu}} - a_{\mu} \left( Q_{\mu} - \sqrt{\frac{a_{\mu}}{2b_{\mu}}} \right)^2 \right.
\nonumber \\
&+& \left. b_{\mu} \left( Q_{\mu} - \sqrt{\frac{a_{\mu}}{2b_{\mu}}} \right)^4
- \frac{1}{2}m_{\mu} {\omega_{\mu}}^2 {Q_{\mu}}^2 \right].
\label{eq3.3}
\end{eqnarray}

\noindent The canonical partition function $Z$ has to be calculated perturbatively upto 1st order in the exponential as

\begin{eqnarray}
Z &=& \int \prod_{\mu}dQ_{\mu} \int \prod_{\mu} dP_{\mu} \exp [-\beta(H_0 + \epsilon H_{1\epsilon})]
\nonumber \\
&=& \prod_{\mu} \int dQ_{\mu} \int dP_{\mu} e^{-\beta H_0} (1 - \epsilon \beta H_{1\epsilon})
\nonumber \\
&=& Z_0 + \epsilon Z_1,
\label{eq3.4}
\end{eqnarray}

\noindent where

\begin{eqnarray}
Z_0 &=& \prod_{\mu} \left[  \int dQ_{\mu} \exp \left\{ -\frac{\beta}{2} m_{\mu} {\omega_{\mu}}^2 {Q_{\mu}}^2 \right\} \right.
\nonumber \\
&\times& \left. \int dP_{\mu} \exp \left\{ -\frac{\beta}{2m_{\mu}} {P_{\mu}}^2 \right\} \right]
\nonumber \\
&=& \prod_{\mu} \frac{2\pi}{\beta \omega_{\mu}}
\label{eq3.5}
\end{eqnarray}

\noindent by simple Gaussian integration and

\begin{eqnarray}
Z_1 &=& -\beta \prod_{\mu} \int dQ_{\mu} \exp \left\{ -\frac{\beta}{2} m_{\mu} {\omega_{\mu}}^2 {Q_{\mu}}^2 \right\}
\nonumber \\
&\times& \int dP_{\mu} \exp \left\{ -\frac{\beta}{2m_{\mu}} {P_{\mu}}^2 \right\}
\nonumber \\
&\times& \sum_{\nu} \left[\frac{{a_{\nu}}^2}{2b_{\nu}} - a_{\nu} \left( Q_{\nu} - \sqrt{\frac{a_{\nu}}{2b_{\nu}}} \right)^2 \right.
\nonumber \\
&+& \left. b_{\nu} \left( Q_{\nu} - \sqrt{\frac{a_{\nu}}{2b_{\nu}}} \right)^4
- \frac{1}{2}m_{\nu} {\omega_{\nu}}^2 {Q_{\nu}}^2 \right].
\label{eq3.6}
\end{eqnarray}

\noindent Realizing that only the even powers from the square brackets enclosed by $\sum_{\nu}$ survive the Gaussian integrations, $Z_1$ takes the form

\begin{eqnarray}
Z_1 &=& \sum_{\nu} \frac{\pi}{\omega_{\nu}}\left[ -\frac{{a_{\nu}}^2}{2b_{\nu}} + \frac{{\Omega_{\nu}}^2}{\beta {\omega_{\nu}}^2}
- \frac{6b_{\nu}}{{\beta}^2 {m_{\nu}}^2 {\omega_{\nu}}^4} \right]
\nonumber \\
&\times& \prod_{\mu \neq \nu} \frac{2\pi}{\beta \omega_{\mu}}
\label{eq3.7}
\end{eqnarray}

\noindent where Eq.(\ref{eq2.10}) has been used in one of the terms. Accordingly, a quantity that we shall require is $Z_1 / Z_0$, and that evaluates to

\begin{equation}
\frac{Z_1}{Z_0} = \sum_{\nu} \frac{\beta}{2}\left[ -\frac{{a_{\nu}}^2}{2b_{\nu}} + \frac{{\Omega_{\nu}}^2}{\beta {\omega_{\nu}}^2}
- \frac{6b_{\nu}}{{\beta}^2 {m_{\nu}}^2 {\omega_{\nu}}^4} \right].
\label{eq3.8}
\end{equation}

\noindent Also, the reciprocal of $Z$, which can be expanded upto first order in $\epsilon$ as

\begin{equation}
\frac{1}{Z} = \frac{1}{Z_0} \left( 1 - \epsilon \frac{Z_1}{Z_0} \right)
\label{eq3.9}
\end{equation}

\noindent will be necessary for calculating the averages. We shall require Eqs.(\ref{eq3.6}) to (\ref{eq3.8}) when we calculate two-time noise correlation in the third subsection.

\subsection{2. The first moment}

\noindent  For calculating the average of the noise defined in Eq.(\ref{eq2.24}),
we shall keep terms upto order $O(\lambda\epsilon)$. Thus

\begin{eqnarray}
\langle \Gamma(t) \rangle &=& \frac{1}{Z} \prod_{\mu} \int dQ_{\mu} dP_{\mu}
\Gamma(\{ Q_{\mu} \}, \{ P_{\mu} \}, t) e^{-\beta(H_0 + \epsilon H_{1\epsilon})}
\nonumber \\
&=& \frac{1}{Z} \prod_{\mu} \int dQ_{\mu} dP_{\mu}
[-\lambda \Gamma_{\lambda}(t) - \lambda\epsilon \Gamma_{\lambda\epsilon}(t)]
\nonumber \\
&\times& e^{-\beta H_0} (1 - \epsilon \beta H_{1\epsilon})
\nonumber \\
&=& -\lambda \langle \Gamma (t) \rangle_{\lambda} - \lambda \epsilon \langle \Gamma (t) \rangle_{\lambda\epsilon}
+ O(\lambda {\epsilon}^2)
\label{eq3.10}
\end{eqnarray}

\noindent where, according to Eq.(\ref{eq3.9}), the averages upto the orders $O(\lambda)$ and $O(\lambda\epsilon)$ have to be calculated with $1/Z_0$, i.e.,

\begin{equation}
\langle \Gamma (t) \rangle_{\lambda} = \frac{1}{Z_0} \prod_{\mu} \int dQ_{\mu} \int dP_{\mu} e^{-\beta H_0} \Gamma_{\lambda} (t)
\label{eq3.11}
\end{equation}

\noindent and

\begin{eqnarray}
\langle \Gamma (t) \rangle_{\lambda\epsilon} &=&
\frac{1}{Z_0} \prod_{\mu} \int dQ_{\mu} \int dP_{\mu} e^{-\beta H_0}
\nonumber \\
&\times& [ \Gamma_{\lambda\epsilon} (t) + \beta H_{1\epsilon} \Gamma_{\lambda}(t)].
\label{eq3.12}
\end{eqnarray}

\noindent From Eqs.(\ref{eq2.25}) and (\ref{eq3.2}) it is clear that the integrals of Eq.(\ref{eq3.11}) become zero due to the presence of odd functions of the bath variables in the integrand. These integrals basically constitute an average of that part of the noise that arises from the harmonic part of the reservoir-Hamiltonian and the average is being taken over the ``unperturbed", i.e., harmonic part of the Hamiltonian as prepared initially.

\noindent The evaluation of the expression on the right hand side of Eq.(\ref{eq3.12}) needs much more care and hence we split it up as

\begin{equation}
\langle \Gamma (t) \rangle_{\lambda\epsilon} = I_1^{(1)} + I_2^{(1)}
\label{eq3.13}
\end{equation}

\noindent such that (here in the superscript we write $(1)$ to denote that these integrals are related to the {\it first} moment)

\begin{eqnarray}
I_1^{(1)} &=& \frac{1}{Z_0} \prod_{\mu} \int dQ_{\mu} \int dP_{\mu} e^{-\beta H_0} \Gamma_{\lambda\epsilon} (t)
\label{eq3.14} \\
I_2^{(1)} &=& \frac{1}{Z_0} \prod_{\mu} \int dQ_{\mu} \int dP_{\mu} e^{-\beta H_0} \beta H_{1\epsilon} \Gamma_{\lambda} (t).
\label{eq3.15}
\end{eqnarray}

\noindent The first integral above, i.e. $I_1^{(1)}$, is the average (again taken over the unperturbed part of the Hamiltonian prepared from the initial values of the bath variables) of that part of the noise which arises form the perturbation-part of the reservoir-Hamiltonian. The integral $I_2^{(1)}$ on the other hand, gives the correlation between the initial nonlinear-part, or the perturbation-part, of the Hamiltonian (i.e., the $H_{1\epsilon}$ part of the integrand) and the unperturbed noise -- again the average being taken over the harmonic or unperturbed-part of the initial Hamiltonian.

\noindent Using the expression for $\Gamma_{\lambda\epsilon} (t)$ from Eq.(\ref{eq2.26}) we have

\begin{eqnarray}
I_1^{(1)} &=& \frac{1}{Z_0} \prod_{\mu} \int dQ_{\mu} \int dP_{\mu} e^{-\beta H_0}
\nonumber \\
&\times& \sum_{\nu} C_{\nu}
[K_0 + K_{1s} \sin \omega_{\nu}(t) + K_{1c} \cos \omega_{\nu}(t)
\nonumber \\
&-& K_{2s} \sin 2\omega_{\nu}(t) - K_{2c} \cos 2\omega_{\nu}(t)
\nonumber \\
&-& K_{3s} \sin 3\omega_{\nu}(t) - K_{3c} \cos 3\omega_{\nu}(t)].
\label{eq3.16}
\end{eqnarray}

\noindent Keeping in mind that $H_0$ [see Eq.(\ref{eq3.2})] contains only quadratic powers of $\{Q_{\mu}\}$ and $\{P_{\mu}\}$, a careful survey of the structures of the $K$-coefficients given in Eqs.(\ref{eq2.12})-(\ref{eq2.18}) reveals that only the terms containing $K_{0}$ and $K_{2c}$ survive the above integrations -- rest all evaluate to zero. The $K_{2c}$-integral, in addition, turns out to be zero finally leading to

\begin{equation}
I_1^{(1)} = \sum_{\nu} \frac{3C_{\nu}}{\beta {m_{\nu}}^2 {\omega_{\nu}}^4} \sqrt{\frac{a_{\nu} b_{\nu}}{2}}
\label{eq3.17}
\end{equation}

\noindent where, for future reference, we note the presence of $a_{\nu}$ in the numerator.

\noindent The expression for $I_2^{(1)}$ as defined in Eq.(\ref{eq3.15}), with the use of Eqs.(\ref{eq2.25}), (\ref{eq3.2}) and (\ref{eq3.3}), takes the form

\begin{eqnarray}
I_2^{(1)} &=& \frac{\beta}{Z_0} \prod_{\mu} \int dQ_{\mu}
\exp \left[ -\frac{\beta}{2}m_{\mu} {\omega_{\mu}}^2 {Q_{\mu}}^2 \right]
\nonumber \\
&\times& \int dP_{\mu} \exp \left[ -\frac{\beta{P_{\mu}}^2}{2m_{\mu}} \right]
\nonumber \\
&\times& \sum_{\nu} \left[\frac{{a_{\nu}}^2}{2b_{\nu}} - a_{\nu} \left( Q_{\nu} - \sqrt{\frac{a_{\nu}}{2b_{\nu}}} \right)^2 \right.
\nonumber \\
&+& \left. b_{\nu} \left( Q_{\nu} - \sqrt{\frac{a_{\nu}}{2b_{\nu}}} \right)^4
- \frac{1}{2}m_{\nu} {\omega_{\nu}}^2 {Q_{\nu}}^2 \right]
\nonumber \\
&\times& \sum_{\sigma} C_{\sigma} \left[ Q_{\sigma} \cos \omega_{\sigma}t
+ \frac{P_{\sigma}}{m_{\sigma} \omega_{\sigma}} \sin \omega_{\sigma}t \right]
\label{eq3.18}
\end{eqnarray}

\noindent The evaluation of this integral proceeds in similar lines as $I_1^{(1)}$ thus reducing it to

\begin{equation}
I_2^{(1)} = \sum_{\nu} \frac{6C_{\nu}}{\beta {m_{\nu}}^2 {\omega_{\nu}}^4} \sqrt{2a_{\nu} b_{\nu}}
\cos \omega_{\nu} t
\label{eq3.19}
\end{equation}

\noindent Finally, combining Eqs.(\ref{eq3.10}), (\ref{eq3.13}), (\ref{eq3.17}) and (\ref{eq3.19}), along with the fact that Eq.(\ref{eq3.11}) evaluates to zero, we have the expression for the first moment of the noise upto order $O(\lambda \epsilon)$ as

\begin{equation}
\langle \Gamma(t) \rangle = -\lambda\epsilon
\sum_{\nu} \frac{3}{2\beta} \frac{C_{\nu}}{{m_{\nu}}^2 {\omega_{\nu}}^4}
\sqrt{2a_{\nu} b_{\nu}} (1 - 12\cos \omega_{\nu} t).
\label{eq3.20}
\end{equation}

\noindent Several points warrant emphasis here. First, if $\epsilon = 0$ -- i.e., the nonlinear part of the reservoir-Hamiltonian in Eq.(\ref{eq1.4}) is put off -- then the first moment of the noise vanishes, as is usually the case with a harmonic reservoir-Hamiltonian. Second -- and contrary to our usual experience -- we have here the first moment of the noise as evaluating to some non-zero quantity not only because of the presence of the nonlinearity in the Hamiltonian of Eq.(\ref{eq1.4}) but also due to an asymmetry in the reservoir potential inflicted by a shift of origin to the bottom of the left well at the very commencement of the above treatment. This term possesses a linear dependence on the temperature at which the initial canonical ensemble was prepared. We observe from the time-dependence of this intrinsic bias of the noise, that the magnitude of the correction term is bounded -- which seems natural as each bath mode, though nonlinear, is pinned to one of the minima of their respective double-well potentials. We shall see briefly in the next Section that when we work with a symmetric nonlinear (quartic) reservoir potential, the first moment of the noise {\it is} zero, because in that case the appearance of $a_{\nu}$ on the right hand side of Eq.(\ref{eq3.20}) is obviated.

\subsection{2. The second moment and the FDR}

\noindent We shall calculate the two-time noise correlation upto third order in perturbation, which precisely turns out to be of the order $O(\lambda^2 \epsilon)$. We shall see that the commonly known form of the FDR in context of system-reservoir theory, viz., the relation that equates the two-time noise correlation with the damping, will come out naturally in the second order of the perturbation parameter $\lambda$. Therefore, our objective is to go one order beyond that to see what corrections to the FDR emerge as a result of nonlinearity in the reservoir oscillators. From Eq.(\ref{eq2.24}) we see that the two-time noise correlation can be written as

\begin{eqnarray}
\langle \Gamma(t) \Gamma(t') \rangle
&=& \langle \{ \lambda \Gamma_{\lambda}(t) + \lambda\epsilon \Gamma_{\lambda\epsilon}(t) \}
\nonumber \\
&\times& \{ \lambda \Gamma_{\lambda}(t') + \lambda\epsilon \Gamma_{\lambda\epsilon}(t') \} \rangle
\nonumber \\
&=& \lambda^2 \langle \Gamma(t) \Gamma(t') \rangle_{\lambda^2}
+ \lambda^2 \epsilon \langle \Gamma(t) \Gamma(t') \rangle_{\lambda^2 \epsilon}
\nonumber \\
&+& O(\lambda^2 \epsilon^2)
\label{eq3.21}
\end{eqnarray}

\noindent where (writing a superscript $(2)$ in naming the integrals involved in calculation of the {\it second} moment) we have

\begin{eqnarray}
\langle \Gamma(t) \Gamma(t') \rangle_{\lambda^2} &=& I_0^{(2)}
\nonumber \\
&=& \frac{1}{Z} \prod_{\mu} \int dQ_{\mu} \int dP_{\mu}
\nonumber \\
&&\times e^{-\beta H_0}
[\Gamma_\lambda(t) \Gamma_\lambda(t')]
\label{eq3.22}
\end{eqnarray}

\noindent and

\begin{equation}
\langle \Gamma(t) \Gamma(t') \rangle_{\lambda^2 \epsilon}
= I_1^{(2)} + I_2^{(2)} + I_3^{(2)}
\label{eq3.23}
\end{equation}

\noindent with

\begin{eqnarray}
I_1^{(2)} &=& \frac{1}{Z} \prod_{\mu} \int dQ_{\mu} \int dP_{\mu} e^{-\beta H_0}
\Gamma_{\lambda} (t) \Gamma_{\lambda\epsilon} (t')
\label{eq3.24} \\
I_2^{(2)} &=& \frac{1}{Z} \prod_{\mu} \int dQ_{\mu} \int dP_{\mu} e^{-\beta H_0}
\Gamma_{\lambda} (t') \Gamma_{\lambda\epsilon} (t)
\label{eq3.25} \\
I_3^{(2)} &=& -\frac{\beta}{Z} \prod_{\mu} \int dQ_{\mu} \int dP_{\mu} e^{-\beta H_0}
\nonumber \\
&&\times H_{1\epsilon} \Gamma_{\lambda}(t) \Gamma_{\lambda}(t')
\label{eq3.26}
\end{eqnarray}

\noindent and from a $(t \leftrightarrow t')$ symmetry in the structures of Eqs.(\ref{eq3.24}) and (\ref{eq3.25}), it is clear that

\begin{equation}
I_1^{(2)} = I_2^{(2)}.
\label{eq3.27}
\end{equation}

\noindent Written explicitly, the right hand side of eq.(\ref{eq3.22}) takes the form

\begin{eqnarray}
I_0^{(2)} &=& \frac{1}{Z} \prod_{\mu} \int dQ_{\mu}
\exp \left[ -\frac{\beta}{2}m_{\mu} {\omega_{\mu}}^2 {Q_{\mu}}^2 \right]
\nonumber \\
&\times& \int dP_{\mu} \exp \left[ -\frac{\beta{P_{\mu}}^2}{2m_{\mu}} \right]
\nonumber \\
&\times& \sum_{\nu} C_{\nu} \left[ Q_{\nu} \cos \omega_{\nu}t
+ \frac{P_{\nu}}{m_{\nu} \omega_{\nu}} \sin \omega_{\nu}t \right]
\nonumber \\
&\times& \sum_{\sigma} C_{\sigma} \left[ Q_{\sigma} \cos \omega_{\sigma}t
+ \frac{P_{\sigma}}{m_{\sigma} \omega_{\sigma}} \sin \omega_{\sigma}t \right].
\label{eq3.28}
\end{eqnarray}

\noindent Since from Eq.(\ref{eq3.2}) we have only quadratic powers of $\{Q_{\mu}\}$ and $\{P_{\mu}\}$ in $H_0$, we see that all terms with $\nu \neq \sigma$ in the above integral become zero. After some algebra this leads to

\begin{equation}
I_0^{(2)} = \frac{1}{Z} \sum_{\nu} \frac{2\pi{C_{\nu}}^2}{\beta^2 m_{\nu}{\omega_{\nu}}^3} \cos \omega_{\nu} (t - t') \prod_{\mu \neq \nu} \frac{2\pi}{\beta \omega_{\mu}}
\label{eq3.29}
\end{equation}

\noindent and finally, with the use of Eqs.(\ref{eq3.5}) and (\ref{eq3.9}), this reduces to

\begin{equation}
I_0^{(2)} = \frac{1}{\beta} \sum_{\nu} \frac{{C_{\nu}}^2}{m_{\nu} {\omega_{\nu}}^2} \cos \omega_{\nu}(t - t')
- \epsilon J_0^{(2)}
\label{eq3.30}
\end{equation}

\noindent where one recognizes the first term on the right hand side as $(k_BT)-$times the expression for damping given in Eq.(\ref{eq2.22}) thus leading to the standard FDR at order $O(\lambda^2)$, and the second term is one of order $O(\lambda^2 \epsilon)$ having the form

\begin{equation}
J_0^{(2)} = \frac{Z_1}{{Z_0}^2} \sum_{\nu} \frac{2\pi{C_{\nu}}^2}{\beta^2 m_{\nu}{\omega_{\nu}}^3} \cos \omega_{\mu} (t - t') \prod_{\mu \neq \nu} \frac{2\pi}{\beta \omega_{\mu}}
\label{eq3.31}
\end{equation}

\noindent With the forms of $Z_0$ and $Z_1 / Z_0$ given respectively in Eqs.(\ref{eq3.5}) and (\ref{eq3.8}), we finally have for the $\lambda^2$ contribution of the two-time noise correlation as

\begin{eqnarray}
\lambda^2 I_0^{(2)} &=& \frac{\gamma}{\beta}
\nonumber \\
&+& \epsilon \sum_{\nu} \gamma
\left[ \frac{{a_{\nu}}^2}{4b_{\nu}} - \frac{{\Omega_{\nu}}^2}{2\beta {\omega_{\nu}}^2}
+ \frac{3b_{\nu}}{{\beta}^2 {m_{\nu}}^2 {\omega_{\nu}}^4}  \right]
\label{eq3.32}
\end{eqnarray}

\noindent which completes the evaluation of the first term on the right hand side of Eq.(\ref{eq3.21}). Here it has to be remembered that $\gamma$ itself is of the order $O(\lambda^2)$.

\noindent Going back to Eq.(\ref{eq3.24}), the right hand side of that equation can be written as

\begin{eqnarray}
I_1^{(2)} &=& \frac{1}{Z} \prod_{\mu} \int dQ_{\mu}
\exp \left[ -\frac{\beta}{2}m_{\mu} {\omega_{\mu}}^2 {Q_{\mu}}^2 \right]
\nonumber \\
&\times& \int dP_{\mu} \exp \left[ -\frac{\beta{P_{\mu}}^2}{2m_{\mu}} \right]
\nonumber \\
&\times& \sum_{\nu} C_{\nu} \left[ Q_{\nu} \cos \omega_{\nu}t
+ \frac{P_{\nu}}{m_{\nu} \omega_{\nu}} \sin \omega_{\nu}t \right]
\nonumber \\
&\times& \sum_{\sigma} C_{\sigma}
[K_0 + K_{1s} \sin \omega_{\sigma}(t) + K_{1c} \cos \omega_{\sigma}(t)
\nonumber \\
&-& K_{2s} \sin 2\omega_{\sigma}(t) - K_{2c} \cos 2\omega_{\sigma}(t)
\nonumber \\
&-& K_{3s} \sin 3\omega_{\sigma}(t) - K_{3c} \cos 3\omega_{\sigma}(t)]
\label{eq3.33}
\end{eqnarray}

\noindent where Eqs.(\ref{eq2.25}), (\ref{eq2.26}), (\ref{eq3.2}) and (\ref{eq3.3}) have been used. The evaluation of this integral is cumbersome but proceeds along similar lines as the previous integrals. A careful survey of the right hand sides of Eqs.(\ref{eq2.12}) to (\ref{eq2.18}) reveals that only the integrals containing $K_{1s}$, $K_{1c}$, $K_{3s}$ and $K_{3c}$ need to be evaluated, and rest all are odd functions and hence zero. The $K_{1s}$ integral evaluates to

\begin{eqnarray}
[K_{1s}~{\rm integral}] &=& \sum_{\nu} \frac{{C_{\nu}}^2}{4 \beta m_{\nu} {\omega_{\nu}}^4}
\left[ {\Omega_{\nu}}^2 - \frac{12 b_{\nu}}{\beta {(m_{\nu} \omega_{\nu})}^2} \right]
\nonumber \\
&\times& \sin \omega_{\nu}t \sin \omega_{\nu}t'
\label{eq3.34}
\end{eqnarray}

\noindent while the $K_{1c}$ integral evaluates to

\begin{eqnarray}
[K_{1c}~{\rm integral}] &=& \sum_{\nu} \frac{{C_{\nu}}^2}{4 \beta m_{\nu} {\omega_{\nu}}^4}
\left[ {\Omega_{\nu}}^2 - \frac{12 b_{\nu}}{\beta {(m_{\nu} \omega_{\nu})}^2} \right]
\nonumber \\
&\times& \cos \omega_{\nu}t \cos \omega_{\nu}t'.
\label{eq3.35}
\end{eqnarray}

\noindent The integrals arising from the $K_{3s}$ and $K_{3c}$ terms of Eq.(\ref{eq3.33}) separately become zero. Therefore, adding the contributions from Eqs.(\ref{eq3.34}) and (\ref{eq3.35}) along with the equality described in Eq.(\ref{eq3.27}), we have

\begin{eqnarray}
I_1^{(2)} + I_2^{(2)} &=& \frac{1}{\beta} \sum_{\nu} \frac{{C_{\nu}}^2}{m_{\nu} {\omega_{\nu}}^2}
\left[ \frac{{\Omega_{\nu}}^2}{2{\omega_{\nu}}^2} - \frac{6 b_{\nu}}{\beta {(m_{\nu} {\omega_{\nu}}^2)}^2} \right]
\nonumber \\
&\times& \cos \omega_{\nu} (t - t').
\label{eq3.36}
\end{eqnarray}

\noindent The final integral to be evaluated is $I_3^{(2)}$ defined in Eq.(\ref{eq3.26}) which, written out explicitly with the help of Eqs.(\ref{eq2.25}), (\ref{eq3.2}) and (\ref{eq3.3}), takes the form

\begin{eqnarray}
I_3^{(2)} &=& -\frac{\beta}{Z} \prod_{\mu} \int dQ_{\mu}
\exp \left[ -\frac{\beta}{2}m_{\mu} {\omega_{\mu}}^2 {Q_{\mu}}^2 \right]
\nonumber \\
&\times& \int dP_{\mu} \exp \left[ -\frac{\beta{P_{\mu}}^2}{2m_{\mu}} \right]
\nonumber \\
&\times& \sum_{\nu} \left[\frac{{a_{\nu}}^2}{2b_{\nu}} - a_{\nu} \left( Q_{\nu} - \sqrt{\frac{a_{\nu}}{2b_{\nu}}} \right)^2 \right.
\nonumber \\
&+& \left. b_{\nu} \left( Q_{\nu} - \sqrt{\frac{a_{\nu}}{2b_{\nu}}} \right)^4
- \frac{1}{2}m_{\nu} {\omega_{\nu}}^2 {Q_{\nu}}^2 \right]
\nonumber \\
&\times& \sum_{\alpha} C_{\alpha} \left[ Q_{\alpha} \cos \omega_{\alpha}t
+ \frac{P_{\alpha}}{m_{\alpha} \omega_{\alpha}} \sin \omega_{\alpha}t \right]
\nonumber \\
&\times& \sum_{\sigma} C_{\sigma} \left[ Q_{\sigma} \cos \omega_{\sigma}t
+ \frac{P_{\sigma}}{m_{\sigma} \omega_{\sigma}} \sin \omega_{\sigma}t \right].
\label{eq3.37}
\end{eqnarray}

\noindent Three pairs of terms give nonzero contributions. The first pair of terms consists of the combinations $Q_{\alpha}Q_{\sigma}\delta_{\sigma\alpha}$ and $P_{\alpha}P_{\sigma}\delta_{\sigma\alpha}$. They evaluate to

\begin{equation}
[{\rm 1st~ pair}] = -\sum_{\sigma} \sum_{\nu} \frac{{a_{\sigma}}^2}{4b_{\sigma}}
\frac{{C_{\nu}}^2}{m_{\nu} {\omega_{\nu}}^2} \cos \omega_{\nu}(t - t').
\label{eq3.38}
\end{equation}

\noindent Combinations constituting the second pair of terms giving nonzero values of the integrals in Eq.(\ref{eq3.37}) are ${Q_{\nu}}^2.Q_{\alpha}Q_{\sigma}\delta_{\sigma\alpha}$ and ${Q_{\nu}}^2.P_{\alpha}P_{\sigma}\delta_{\sigma\alpha}$ and they evaluate to

\begin{equation}
[{\rm 2nd~ pair}] = \frac{1}{\beta} \sum_{\sigma} \sum_{\nu} \frac{{\Omega_{\sigma}}^2}{2{\omega_{\sigma}}^2}
\frac{{C_{\nu}}^2}{m_{\nu} {\omega_{\nu}}^2} \cos \omega_{\nu}(t - t').
\label{eq3.39}
\end{equation}

\noindent The third pair consists of terms like ${Q_{\nu}}^4.Q_{\alpha}Q_{\sigma}\delta_{\sigma\alpha}$ and ${Q_{\nu}}^4.P_{\alpha}P_{\sigma}\delta_{\sigma\alpha}$ and they yield

\begin{equation}
[{\rm 3rd~ pair}] = -\frac{1}{\beta^2} \sum_{\sigma} \sum_{\nu} \frac{3b_{\sigma}}{(m_{\sigma} {\omega_{\sigma}}^2)^2}
\frac{{C_{\nu}}^2}{m_{\nu} {\omega_{\nu}}^2} \cos \omega_{\nu}(t - t').
\label{eq3.40}
\end{equation}

\noindent Combining the right hand sides of Eqs.(\ref{eq3.38}) - (\ref{eq3.40}) we have

\begin{eqnarray}
I_3^{(2)} &=& -\sum_{\sigma} \left[ \frac{{a_{\sigma}}^2}{4b_{\sigma}}
- \frac{1}{\beta} \frac{{\Omega_{\sigma}}^2}{2{\omega_{\sigma}}^2}
+ \frac{1}{\beta^2} \frac{3b_{\sigma}}{(m_{\sigma} {\omega_{\sigma}}^2)^2} \right]
\nonumber \\
&\times& \sum_{\nu} \frac{{C_{\nu}}^2}{m_{\nu} {\omega_{\nu}}^2} \cos \omega_{\nu}(t - t').
\label{eq3.41}
\end{eqnarray}

\noindent Therefore, the right hand side of Eq.(\ref{eq3.23}) is obtained by summing those of Eqs.(\ref{eq3.36}) and (\ref{eq3.41}). Eventually, by adding to this quantity the $O(\lambda^2\epsilon)$ term appearing in Eq.(\ref{eq3.32}), the overall term of order $O(\lambda^2\epsilon)$ emerging from the expression of two-time noise correlation -- viz., Eq.(\ref{eq3.21}) -- is obtained. In doing so we find that the right hand side of Eq.(\ref{eq3.41}) {\it exactly cancels} the third order term on the right hand side of Eq.(\ref{eq3.32}), because the expression for damping given in Eq.(\ref{eq2.22}) appears in Eq.(\ref{eq3.41}). Thus, the task of evaluating the two-time noise correlation to order $O(\lambda^2 \epsilon)$ has now been accomplished. Recalling that $\gamma$ itself is of order $O(\lambda^2)$ we have

\begin{eqnarray}
\langle \Gamma(t) \Gamma(t') \rangle &=& \frac{\gamma}{\beta}
\nonumber \\
&+& (\lambda^2 \epsilon)
\sum_{\nu} \frac{D_{\nu}{C_{\nu}}^2}{m_{\nu} {\omega_{\nu}}^2} \cos \omega_{\nu} (t - t').
\label{eq3.42}
\end{eqnarray}

\noindent
where, for notational brevity, we have defined $D_{\nu} \equiv D_{\nu}(\beta, \{R_{\nu}\})$ (here $\{R_{\nu}\}$ denotes the set of parameters characterizing the $\nu$-th oscillator of the reservoir)

\begin{eqnarray}
D_{\nu}(\beta, \{R_{\nu}\}) &=& \frac{1}{\beta} \frac{{\Omega_{\nu}}^2}{2{\omega_{\nu}}^2}
- \frac{1}{\beta^2} \frac{6 b_{\nu}}{{(m_{\nu} {\omega_{\nu}}^2)}^2}
\nonumber \\
&=& \frac{1}{\beta} \left( \frac{1}{2} - \frac{2a_{\nu}}{m_{\nu} {\omega_{\nu}}^2} \right)
- \frac{1}{\beta^2} \frac{6 b_{\nu}}{{(m_{\nu} {\omega_{\nu}}^2)}^2}
\nonumber \\
\label{eq3.43}
\end{eqnarray}

\noindent that contains higher powers of $1/\beta$, i.e., $k_BT$. The first term on the right hand side of Eq.(\ref{eq3.42}) is the standard FDR that emerges from coupling of the system with a harmonic bath. Therefore we see that, nonlinearity in the reservoir oscillators bring higher order corrections to the FDR and get manifested through higher powers of $k_BT$. Based on the above results, we now briefly describe the situation for a symmetric quartic bath.

\section{IV. The symmetric quartic bath}

\noindent By a symmetric quartic bath we mean that the reservoir Hamiltonian of Eq.(\ref{eq1.4}) gets simplified to

\begin{equation}
H_R = \sum_{\mu} \left[ \left\{\frac{{p_{\mu}}^2}{2m_{\mu}} + \frac{1}{2}m_{\mu} {\omega_{\mu}}^2 {q_{\mu}}^2 \right\}
+ \epsilon b_{\mu} {q_{\mu}}^4 \right]
\label{eq4.1}
\end{equation}

\noindent which means that the potential $U_{\mu}(q_{\mu}) = \frac{1}{2}m_{\mu} {\omega_{\mu}}^2 {q_{\mu}}^2
+ \epsilon b_{\mu} {q_{\mu}}^4$ is now symmetric about the origin and hence no initial change of origin is required [as was done in Eq.(\ref{eq1.1})] to effect a perturbative treatment. The harmonic (and hence solvable) part and the perturbation part of the Hamiltonian of Eq.(\ref{eq4.1}) are clearly spelt out. The other terms of the total system-reservoir Hamiltonian of Eq.(\ref{eq1.2}), viz. $H_S$, $H_C$ and $H_{CT}$ are kept intact as they were in Eqs.(\ref{eq1.3}), (\ref{eq1.5}) and (\ref{eq1.6}) respectively.

\noindent Our detailed treatment of the double-well case in Sections 1 to 3 makes our task immensely easier for this Section and hence we shall just state the main results which follow quite obviously from the previous ones. First of all, every $a_{\mu}$ containing term will become zero -- because there is no $a_{\mu}$ in the Hamiltonian of Eq.(\ref{eq4.1}) and hence in the full Hamiltonian of Eq.(\ref{eq1.2}) -- and, the frequency $\Omega_{\mu}$ defined in Eq.(\ref{eq2.10}) should also be zero -- because this particular combination of terms containing $\omega_{\mu}$ and $a_{\mu}$ arose due to the addition and a subsequent compensatory subtraction of the harmonic term $\frac{1}{2}m_{\mu} {\omega_{\mu}}^2 {q_{\mu}}^2$ in the reservoir Hamiltonian of Eq.(\ref{eq1.4}) and that technique is of no relevance for this symmetric case. Therefore, our main two results relating to the moments of the noise get much simplified.

\noindent For the first moment Eq.(\ref{eq3.20}) reduces to

\begin{equation}
\langle \Gamma(t) \rangle = 0.
\label{eq4.2}
\end{equation}

\noindent As was commented in the discussion following Eq.(\ref{eq3.20}), the vanishing of the first moment of the noise, in keeping with our usual experience, happens because of the symmetry in the reservoir potential. In the double-well case, due to a shift of origin the noise got a bias on one side which prevented its first moment from becoming zero.

\noindent For the two-point noise correlation, we have from Eqs.(\ref{eq3.42}) and (\ref{eq3.43})

\begin{eqnarray}
\langle \Gamma(t) \Gamma(t') \rangle =\frac{\gamma}{\beta}
+ \frac{(\lambda^2 \epsilon)}{\beta^2}
\sum_{\nu} \frac{6 b_{\nu}{C_{\nu}}^2}{{(m_{\nu} {\omega_{\nu}}^2)}^3} \cos \omega_{\nu} (t - t')
\nonumber \\
\label{eq4.3}
\end{eqnarray}

\noindent where we again notice that higher powers of $k_BT$ remain present; hence this correction to the FDR, arising due to nonlinearity in the reservoir potential, is much more fundamental that go beyond the structure of the potential -- symmetric or asymmetric -- about one of its minima with respect to which the perturbation theory is being built up.


\section{V. Conclusion}

\noindent In this paper we have addressed the classical problem of a general system coupled linearly to a bath of nonlinear oscillators. Nonlinear reservoirs have not come into direct focus from the microscopic viewpoint, where one employs Zwanzig's approach for forming the generalized Langevin equation by eliminating the bath degrees of freedom. Here we extend that program to study the effect of nonlinearities in the bath oscillators through perturbation theory. After explaining in detail the rationale behind the Hamiltonian -- that involves two different perturbation parameters, one for the coupling and the other for the nonlinearity -- in Section I, and subsequently formulating the generalized Langevin equation in Section II, we have gone into detailed calculation for the evaluation of the relevant noise correlations perturbatively in Section III. For double-well bath oscillators, the perturbation theory has been built up about one of the minima, and this construction makes the potential asymmetric about this chosen origin, thus making the first moment of the noise non-zero in second order in perturbation. More significant is the fact that the fluctuation dissipation relation also gets corrected at the third order and this correction involves nonlinear powers of $k_BT$ in addition to the usual FDR (for harmonic baths) that we retrieve at the second order in perturbation. Section IV has been rather brief and has been devoted to the structure of the aforementioned results for a symmetric quartic bath. Due to symmetry the first moment of the noise becomes zero but the corrections to the FDR in higher orders remain.

\noindent The purpose of this paper is to create a groundwork for some futuristic directions relating to fundamental issues where system-reservoir studies with harmonic baths have rightfully remained as a basic and powerful language. For example, in studies pertaining to activated rate processes in classical as well as quantum domains -- studies that have built up for almost a century based on the celebrated Arrhenius rate formula -- it is worthwhile to expect that the correction terms deduced in this work may bring in new results. Nonlinear oscillators are also important in transport processes that describe heat conduction from one reservoir to another through a prescribed channel. In such processes where energy gets localized due to non-equipartition among the nonlinear modes, it is worthwhile looking into whether the correction terms can throw some new light on the various domains of diffusivities which come up in such contexts. To keep the paper absolutely general we have not addressed any particular system, which, of course, remains as another application ground for the results derived.

\acknowledgements

\textbf{Acknowledgements:} One of the authors (CB) is grateful to Professor Ulrich Weiss (University of Stuttgart, Germany) and Professor Michele Campisi (Scuola Normale Superiore, Italy) for their critical comments when he met them in FQMT-2015 held in Prague in August 2015. He is thankful to the Council of Scientific and Industrial Research, India, for financial support. The authors are also grateful to Professor Deb Shankar Ray (IACS, Kolkata, India) for useful discussions.

\end{document}